\documentclass[aps,prl,twocolumn,superscriptaddress]{revtex4-1}

\usepackage{graphicx}
\usepackage{amsmath}
\usepackage{amssymb}
\usepackage{color}

\newcommand{\rs}{\rm \scriptscriptstyle}

\begin{document}

\title{Artificial atoms can do more than atoms: Deterministic single photon\\ subtraction from arbitrary light fields}


\author{Jens Honer}

\affiliation{Institute for Theoretical Physics III, University of Stuttgart, Germany}

\author{R. L\"ow}
\affiliation{5.~Physikalische Institut, University of Stuttgart, Germany}

\author{Hendrik Weimer}
\affiliation{Institute for Theoretical Physics III, University of Stuttgart, Germany}
\affiliation{ Physics Department, Harvard University, Cambridge, Massachusetts 02138, USA}
\affiliation{ITAMP, Harvard-Smithsonian Center for Astrophysics, Cambridge, Massachusetts 02138, USA}

\author{Tilman Pfau}
\affiliation{5.~Physikalische Institut, University of Stuttgart, Germany}

\author{Hans Peter B\"uchler}
\affiliation{Institute for Theoretical Physics III, University of Stuttgart, Germany}

\date{\today}

\begin{abstract}
  We study the interplay of photons interacting with an artificial atom  in the presence of a controlled dephasing. 
  Such artifical atoms consisting of several independent scatterer can exhibit remarkable properties superior to
  single atoms with a prominent example being a superatom based on Rydberg blockade.  We demonstrate that the induced dephasing 
  allows for the controlled absorption of a single photon from  an arbitrary incoming probe field.
  This  unique tool in photon-matter interaction opens a way for building novel quantum devices and several potential applications 
  like a single photon transistor, 
  high fidelity n-photon counters, or for the creation of non-classical states of light by photon subtraction are presented.
\end{abstract}


\maketitle

The prime model for our understanding of resonant light matter interaction
is based on two energy levels of a single atom coupled by a dipole moment to
the electromagnetic radiation.
%
On the other hand,  artificial atoms
\cite{nphys469,Michler22122000,nnphot,nature05461,natureGirvin,nature08134,Neumann30072010},
which are based on two states of a more complex quantum many body system,  can
exhibit properties superior to the conventional single atom. Here, we show that
an artificial superatom made of a large number of scatterers under the
influence of a blockade mechanism exhibits the extraordinary property to
deterministically absorb a single photon of an arbitrary incoming light field.
The main idea is based on an externally induced controlled dephasing for the
excitations. This unique tool in photon-matter interaction opens a way
for building novel quantum devices like a single photon transistor, high
fidelity $n$-photon counters, or for the creation of non-classical states of
light by  photon subtraction
\cite{Ourjoumtsev07042006,PhysRevLett.97.083604,Parigi28092007}.

One way to reduce the physics of a mesoscopic ensemble of $N$
scatterers to two relevant energy levels is to introduce a strong
interaction between the excited states. Such a scenario is realized
for an ensemble of atomic atoms driven resonantly into a Rydberg
level giving rise to the so-called Rydberg blockade phenomena
\cite{ PhysRevLett.93.163001,PhysRevLett.93.063001,PhysRevLett.97.083003,PhysRevLett.99.163601,RevModPhys.82.2313}, but also for semiconductor quantum dots \cite{nnphot}.
In these cases the excited state is the coherent superposition
with a single excitation shared among  the particles providing an enhanced coupling $\sim \sqrt{N}$ 
compared to an individual scatterer \cite{nature06331,nphys1183,PhysRevLett.105.193603}.
%
It is this strong coupling,  which features superatoms
as ideal quantum information processor for photons \cite{nature03064,natureEisaman,nature04315}, while
many  tools currently developed for  single atom-photon 
coupling (see \cite{nphys1805} and references therein) can be directly carried over.
On the other hand,  superatoms  can exhibit  phenomena, which are superior to conventional atoms, e.g., 
as strongly directed single photon source \cite{PhysRevA.78.053816,PhysRevA.66.065403,nnphys844}.
%


\begin{figure}[htbp]
  \includegraphics[width= 1\columnwidth]{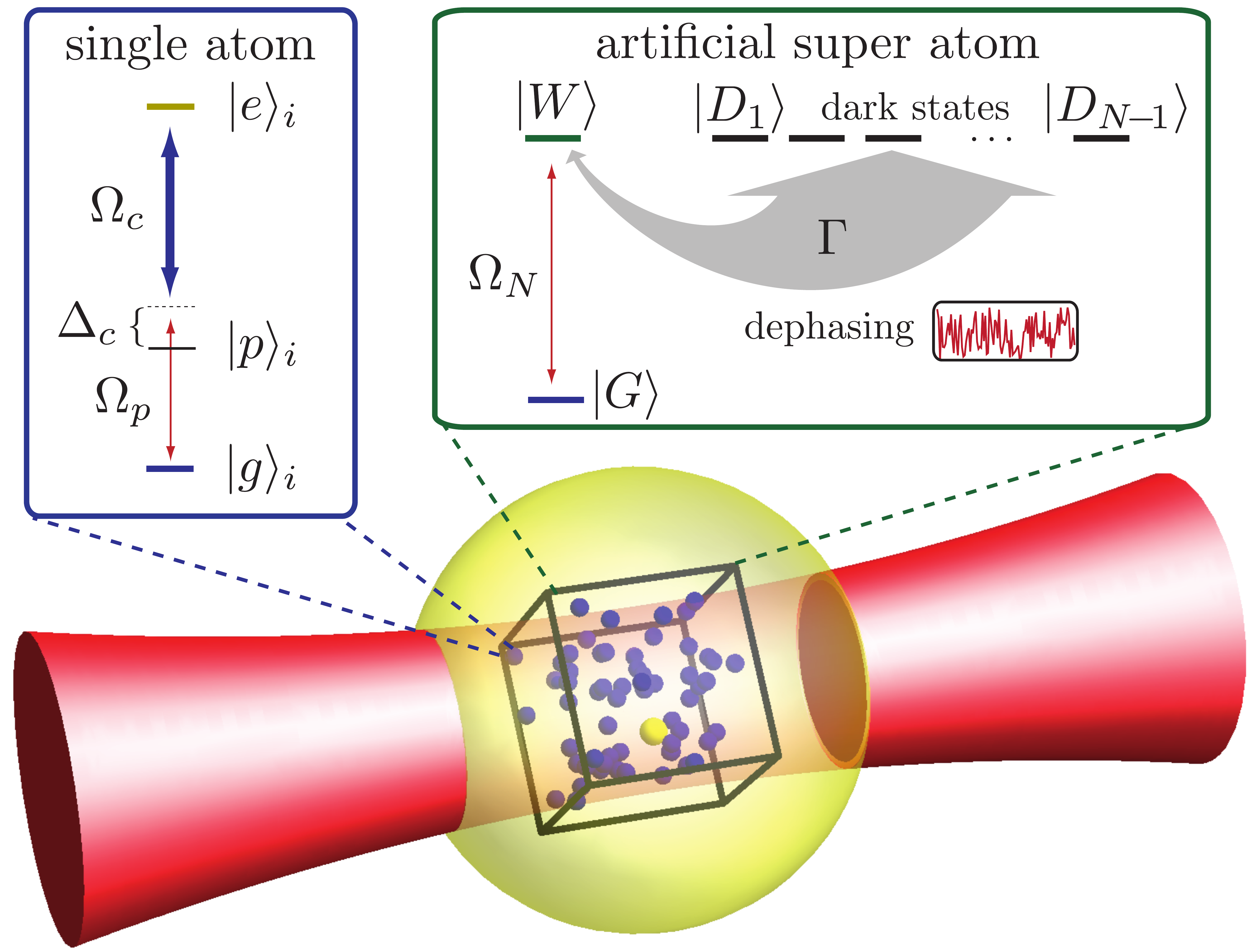}
\caption{Artificial superatom based on Rydberg atoms: for each atom the transition to the Rydberg state takes place via an intermediate
 $p$-level and a strong coupling laser with Rabi frequency $\Omega_{c}$ and detuning $\Delta_{c}$. 
 Within the mode volume of the probe field with Rabi frequency $\Omega_{p}$, there are $N$ atoms within
 the collective blockade radius (yellow sphere) exceeding the cell size.  The system gives rise to a superatom state 
 $|W\rangle$ with a single Rydberg excitation shared among the particles with collective Rabi frequency 
 $\Omega_{N}$. In addition, the system exhibits $N\!-\!1$ dark states 
 $|D_{j}\rangle$, which are coupled to the superatom state by a controlled inhomogeneous dephasing with rate $\Gamma$.}
\label{fig1}
\end{figure}

Here, we show that an artificial superatom  based on Rydberg
blockade provides a high fidelity and saturable single photon
absorber, i.e., for an arbitrary incoming probe field, a single
photon is subtracted and a Rydberg excitation generated, 
while the remaining part of the probe field propagates
through the medium. In contrast, for a single atom the limit to
absorb a single photon with a probability higher than
50\% is only achieved  by stringent 
requirements on the pulse shape or arrival time: a well controlled 
$\pi$-pulse and its analogue on the single photon level  \cite{EPL14007}  
can coherently excite a single atom, while STIRAP (stimulated Raman adiabatic passage)  
demands control on the  arrival time of the probe field. The scheme presented here circumvents these
restrictions and provides a unique tool for quantum information
processing. We demonstrate applications of this setup for the design of a 
single photon transistor, a high fidelity $n$-photon detector, and the creation of non-classical 
states of light by photon subtraction.

The main idea is based on the combination of enhanced light-matter 
coupling in the mesoscopic ensemble with a controlled dephasing 
between the atoms. More specifically, the superatom possesses   
$N\!-\!1$ dark states in addition to the two bright states forming the
 two-level system, see Fig~\ref{fig1}. Using an external dephasing allows
us to couple the collective excited state to these dark states,
and distribute the excited state population over all $N$ levels.
As a consequence,  one obtains an equal population of these states
and the probability to absorb a single photon scales as $N/(N+1)$; it reaches 
unity for large number of scatterers involved. While this general
method can  be applied to arbitrary artifical atoms based on a blockade mechanism, 
we will present the analysis for an ensemble of atoms laser driven into a
Rydberg state.

The setup for the superatom consists of $N$ atoms confined 
into a small trap, which can be excited into a Rydberg state $|e\rangle_{i}$ 
through a two-photon transition: a strong laser 
field couples the Rydberg state $|e\rangle_{i}$ and the intermediate 
level $|p\rangle_{i}$ with Rabi frequency $\Omega_{c}$ and detuning 
$\Delta_{c}\gg \Omega_{c}$,  while the transition  from the  ground 
state  $|g\rangle_{i}$  to the intermediate $|p\rangle_{i}$ level is driven 
by  the probe field with Rabi frequency $\Omega_{p}$, see Fig.~\ref{fig1}. 
In the following, we assume near resonant light for this two-photon transition.
The strong interaction between the Rydberg atoms gives rise to a collective 
blockade radius $r_{\rs B} = (C_{6}/\hbar \sqrt{N} \Omega)^{1/6}$ 
with the two-photon Rabi frequency $\Omega = \Omega_{c}\Omega_{p}/4 \Delta_{c}$ \cite{PhysRevLett.99.163601}.
We are interested in the regime, where all atoms are trapped within the Blockade radius.
Then, only a single Rydberg excitation is possible and the relevant states of the N-body system 
is the state $|G\rangle$ with all atoms in the ground state and the excited states
$|i\rangle$ with the $i^{\rm th}$ atom excited to the Rydberg state. 
In addition, we require that the probe beam is strongly focused with a transverse 
mode area $A$ smaller than the  size of the transverse trapping of the atomic system, see Fig.~\ref{fig1}. 
With a characteristic size of the Blockade radius in the range of $5 {\rm \mu m}$ \cite{PhysRevLett.99.163601},
these conditions can be well satisfied with current techniques for cold atomic gases.

In the `frozen' Rydberg regime, 
the system reduces to a superatom with the ground state $|G\rangle$ and the excited 
$W$-state  $|W\rangle =\sum_{i} |i\rangle/\sqrt{N}$  accounting for the coherent 
superposition of a single Rydberg excitation. The coherent dynamics of this two-level system 
is given by the Hamiltonian 
\begin{equation}
H= \frac{\hbar \Omega_{N}}{2} \Big[\: | W\rangle \langle G| + |G \rangle \langle W|\:\Big]
\end{equation}
describing the coupling between the two bright states  with the collective 
Rabi frequency  $\Omega_{N} = \sqrt{N}\Omega$.

Several mechanisms can lead to decoherence and dephasing of the 
superatom. We distinguish between {\it homogeneous}
and {\it inhomogeneous} decoherence: the first one acts on all atoms in the
same way, e.g., fluctuations in the phase of the driving lasers,
and leads to the dephasing of the superatom in analogy to a single atom. 
In turn, the inhomogeneous  dephasing will influence each atom individually
and gives rise to a fundamental differences between a single atom and the superatom. 

\begin{figure}[h]
 \includegraphics[width= 0.8\columnwidth]{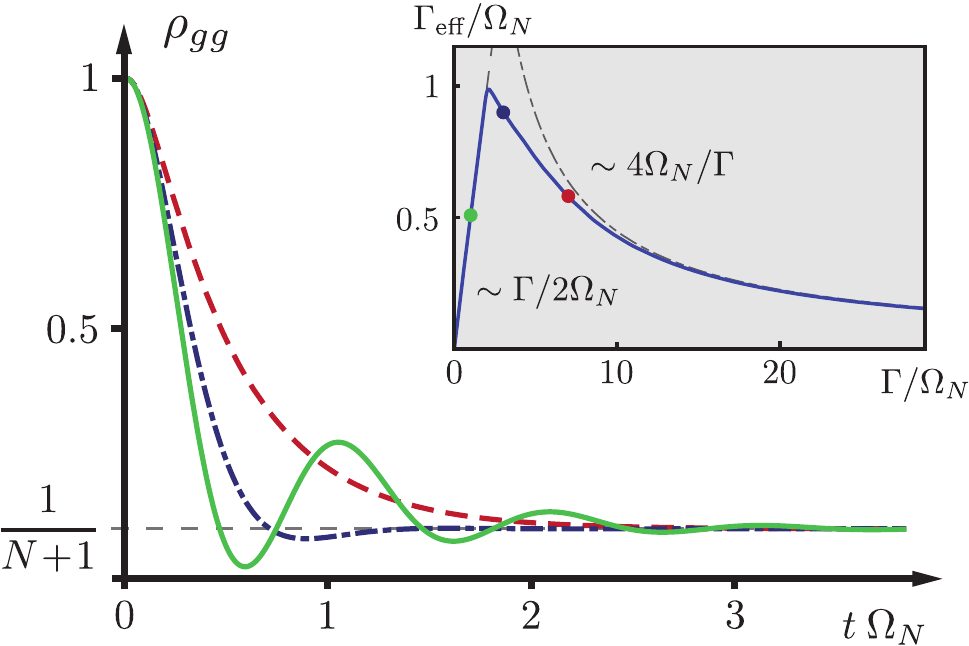}
\caption{Time evolution: the probability $\rho_{gg}$ for the atomic system to remain in the 
   ground state decays exponentially with the rate $\Gamma_{\rs eff}$ to the final value $1/(N+1)$. 
   The numerical integration is performed for $N = 9$ in the overdamped regime for $\Gamma= 7\Omega_{N} $ (red line), 
   as well as in the underdamped regime with $\Gamma=\Omega_{N} $ (green line), and in the crossover $\Omega_{N} = 3 \Gamma$  (blue line).  
   The inset shows the dependence of the effective damping rate $\Gamma_{\rs eff}$ from the collective Rabi frequency $\Omega_{N}$, with $\Gamma_{\rs eff} = \Gamma/2$ for
   weak dephasing, while the crossover from the under- to overdamped regime appears at $\Omega_{n} \sim 3 \Gamma$, the damping rate scales as $\Gamma_{\rs eff} \sim 4 \Omega_{N}^2/\Gamma$.}
\label{fig2}
\end{figure}

An ideal {\it inhomogeneous} dephasing acting on each Rydberg atom
can be microscopically designed by an ac Stark shift $\Delta_{i}(t)$ of the Rydberg level by a speckle 
light field with short range correlations in space  as well as in time \cite{speckle}, i.e., the detuning 
exhibits a spatial correlation $\xi$ comparable to the interparticle distance, and a temporal correlation $\tau$ shorter than  the collective Rabi frequency.  
%

%
Then, the random distribution of the atoms in the `frozen' Rydberg regime gives rise to the 
Hamiltonian accounting for the noise term
\begin{equation}
  H_{\rs noise} =  \sum_{i}  \Delta_{i}(t) \: |i \rangle \langle i|
\end{equation}
with the correlations between the different detunings
$\langle \Delta_{i}(t) \Delta_{j})(t')\rangle =\Gamma \delta_{i j} \delta(t-t')$. 
This term in the Hamitlonian
couples the excited bright state $|W\rangle$ with the additional dark states $|D_{j}\rangle$.
Note,  that an 
alternative microscopic sources for  {\it inhomogeneous dephasing} is obtained by the combination of a 
pulsed coupling laser with optical lattices. In addition,  the motion 
of the atoms beyond the frozen Rydberg approximation is a intrinsic mechanism leading to 
inhomegenous dephasing.
%
%
%
%
Then, the Lindblad master equation for the two bright states
and the $N-1$ dark states becomes
\begin{equation}
   \partial_{t} \rho = -  \frac{i}{\hbar} \left[H, \rho \right] +\Gamma \sum_{i}^{N}  \left(2  c_{i} \rho  c_{i}^{\dag} - c^{\dag}_{i} c_{i} \rho -\rho  c^{\dag}_{i} c_{i} \right)
   \label{inhomogeneousdephasing}
\end{equation}
with the jump operators $c_{i} = | i\rangle\langle i |$.  The dephasing $\Gamma$ gives 
rise to an exponential decay of the coherences
in the system, and the stationary solution for the density matrix reduces to
\begin{equation}
   \rho = \frac{1}{N+1}\left(|G \rangle \langle G| + |W\rangle \langle W| +\sum_{j}^{N-1} |D_{j} \rangle \langle D_{j}| \right).
\end{equation}
The fidelity to absorb a photon  $f=
1- {\rm Tr}(\rho |G\rangle \langle G|)= N/(N+1)$  is strongly enhanced
for large particle numbers and eventually approaches unity for $N \gg 1$.
In contrast,  for a single atom, this reduces to the well known fundamental 
limit that on average we can only absorb a photon with probability $f=1/2$. 
The full dynamics of the system can be obtained by a straightforward time
integration of  the master equation, see Fig.~\ref{fig2}.
For small dephasing $\Gamma< \Omega_{N}$ the system still exhibits
Rabi oscillations, which eventually are damped out with the effective decay rate
$\Gamma_{\rs eff} = \Gamma$, while for
increasing dephasing $\Gamma$ the system turns overdamped with
$\Gamma_{\rs eff} \sim \Omega_{N}^2/\gamma$. 
The shortest time for equilibration is achieved at the crossover from the underdamped
to the overdamped regime.

The validity for the setup to work for a probe pulse containing a few photons
requires, that a single photon is absorbed within the coherence time $\tau$
of the probe pulse, i.e.,   $\Gamma \tau > 1$. Using the single photon Rabi frequency this condition
reduces to  optical thick media $\kappa > 1$ with
\begin{equation}
  \kappa = 2 \pi N \frac{d^2}{\hbar c A} \frac{\omega_{p}}{\Gamma} \frac{\Omega_{c}^{2}}{\Delta_{c}^2},
\end{equation}
the transverse mode area $A$, the dipole transition moment $d$, and the frequency of the
probe field $\omega_{p}$. In turn, the full absorption has to take place on a time scale shorter than the
spontaneous emission from the Rydberg level. These conditions are naturally satisfied for cold Rydberg gases.

\begin{figure}[h]
 \includegraphics[width= 1\columnwidth]{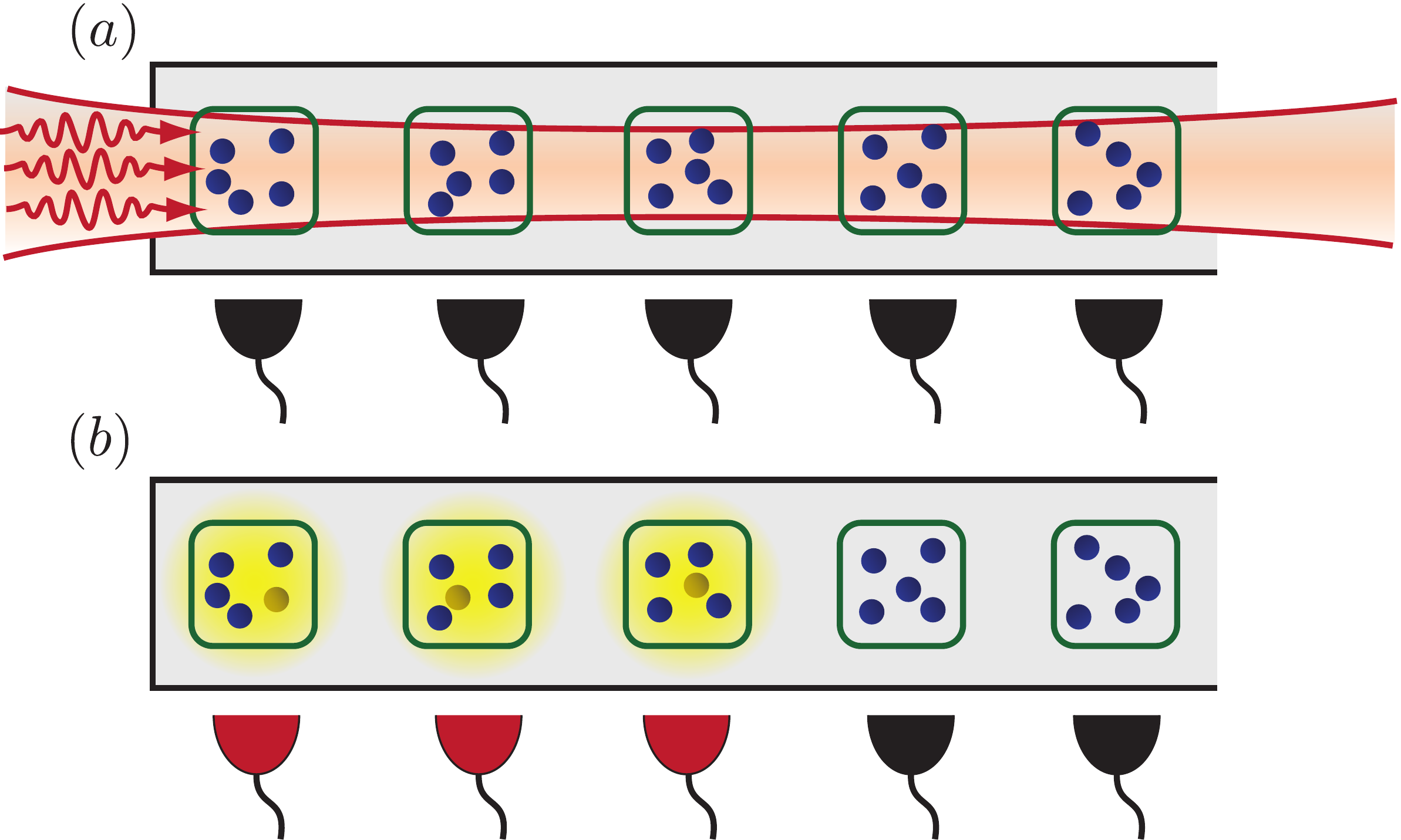}
\caption{Setup for a high fidelity $k$-poton detector: (a) illustration of the incoming probe field with three photons on a chain of several 
   cells, which containing a deterministic single photon absorber. Each cell absorbs  exactly one photon, while the remaining beam 
   propagates to the next cell.
 (b) In the first three cells exactly one photon is absorbed and an atom is excited to the Rydberg state, 
      which in turn is subsequently detected: the position of the last Rydberg excitation provides the number of photons within the beam.}
\label{fig3}
\end{figure}

In combination with a robust detection scheme of the Rydberg excitation, the
setup give rise to a saturable single photon detector with near unity fidelity.  Here, we
envisage the detection of the Rydberg excitation via the combination of electromagnetic 
induced transparency  (EIT)
and Rydberg blockade \cite{PhysRevLett.102.170502,PhysRevLett.102.240502}.
The system is probed by a weak detection beam satisfying the EIT condition, i.e., its frequency is shifted by $\Omega_{c}^2/4 \Delta_{c}$
compared to the probe field. Then, the detection beam passes through the cell without
any  phase shift if  all atoms  are in the ground state. However, as soon as a 
single Rydberg state is excited, the EIT condition is violated and the detection beam acquires a phase shift due to the real part in 
the dielectric response function \cite{RevModPhys.77.633}. The subsequent measurement of the phase shift via a homodyne detection  provides 
a perfect single photon detector. In gate operation language, the setup acts as a classical single photon transistor: a single photon 
in the probe beam, switches several photons in the homodyne detector.

Combining several single photon detectors  into a chain of individually 
addressable cells then opens the unique possibility for a high fidelity $k$-photon detector, see Fig.~\ref{fig3}: 
each cell absorbs only a single photon,  while the probe beam reduced by a photon propagates to the 
next detector. The subsequent measurement of the Rydberg excitation for each cell provides the deterministic 
detection of the number of photons in the probe field.

\begin{figure}[h]
\includegraphics[width= 1\columnwidth]{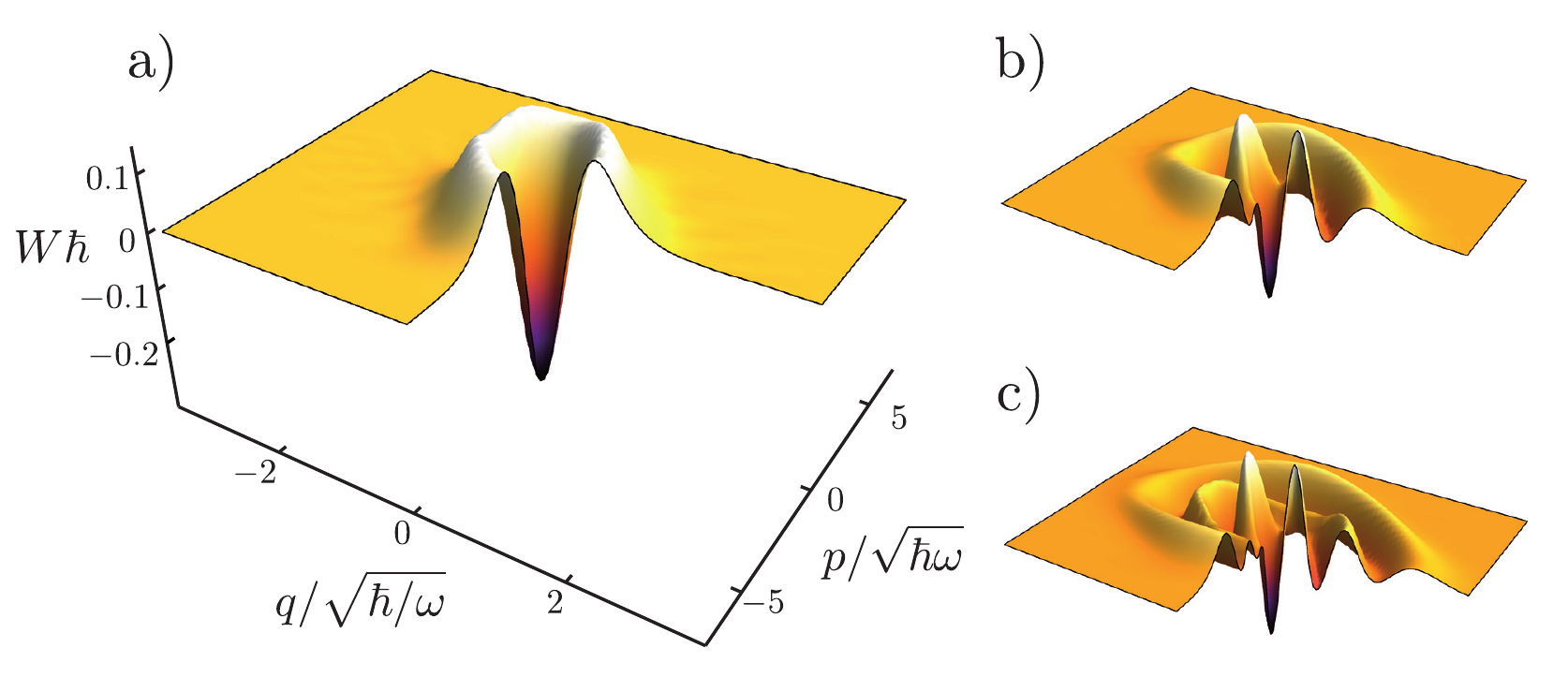}
\caption{Non-classical states of light:  (a) Wigner function of the outcoming photon state after the deterministic absorption of a photon within the
cell.  The incoming state is characterized by a  coherent state  ($\alpha= 0.2$) with subsequent amplitude squeezing  ($w=0.6$). The Wigner function exhibits 
close resemblance to a photonic ``cat'' state. In addition, the Wigner function for the deterministic 3-photon absorption (b), as well as 5-photon 
absorption (c) are shown.}
\label{fig4}
\end{figure}

Finally, the subtraction of photons from a squeezed vacuum field has extensively been discussed for the creation of
non-classical states of light such as cat states \cite{Ourjoumtsev07042006,PhysRevLett.97.083604,Parigi28092007}.
A major drawback of these schemes is a low probability for the  reduction of the incoming beam by a single photon. Here,
our scheme provides an advantage as a photon is subtracted, whenever a photon is present in the squeezed vacuum.
 The influence on the incoming probe field of the single photon absorption  
is well accounted for in a master equation approach. The single photon detector is described by 
two states: $|G \rangle$ for the atoms in the ground state and $|E \rangle$ for the state of the 
cell with a single Rydberg excitation detected. Then, the master equation describing the dynamics
of the photon absorption and detection reduces to
\begin{equation}
   \partial_{t} \rho = \Gamma_{\rs eff} \left(2  c \rho  c^{\dag} - c^{\dag} c \rho -\rho  c^{\dag} c \right) \label{ME2},
\end{equation}
where  the jump operator $ c = a \: |E\rangle \langle G|$ contains the photon annihilation operator 
$a$  and describes the transition to the state with a photon detected, while $\Gamma_{\rs eff}$ accounts for the time 
scale for the single photon absorption, see Fig.~\ref{fig2}. Note, that the jump operator satisfies $c^2 = 0$ 
corresponding to the fact that a single photon saturates the absorber.
The master equation can be analytically solved for an initial state 
$\rho= |G\rangle \langle G| \otimes \rho_{\rs inc}$ with $\rho_{\rs inc} = \sum_{n m} \rho_{n m} |n \rangle \langle m|$
the density matrix for the incoming probe field. The absorption of the photon takes place on the 
time scale $\Gamma_{\rs eff}$ and for $t \gg 1/\Gamma_{\rs eff}$ the density matrix approaches 
 the stationary solution 
\begin{equation}
\rho \rightarrow p_{\rs vac} |G\rangle \langle G| \otimes \rho_{\rs vac}  + (1-p_{\rs vac}) |E\rangle \langle E | \otimes \rho_{\rs out}
\end{equation}
with $p_{\rs vac}$ the probability for the initial photon state to be in the vacuum $\rho_{\rs vac} = |0\rangle \langle 0|$. The density matrix describing the 
outgoing photon field takes the form
\begin{equation}
 \rho_{\rs out}= \frac{1}{1-p_{\rs vac}}\sum_{ n ,m=1} \frac{2}{n+m}  \sqrt{n m} \rho_{n m}  | n-1\rangle \langle m-1|.
 \label{outgoingbeam}
\end{equation}
The Wigner function for this density matrix with an incoming squeezed state is shown in Fig.~\ref{fig4},
and exhibits the characteristic features of a cat state created by photon subtraction \cite{Ourjoumtsev07042006,PhysRevLett.97.083604,Parigi28092007}.
Consequently, with probabilty $p_{\rs vac}$ no photon is present in the incoming beam, and the
outgoing beam remains in the vacuum state $\rho_{\rs vac} = |0\rangle \langle 0|$, while
with probability $1-p_{\rs vac}$ a photon is absorbed from the incoming beam.
Note, that this probability represents a 
fundamental limit on the fidelity to generate  non-classical states of light by  single photons subtraction.

To summarize we have identified a  novel feature of artificial atoms
based on a strong excited state blockade. Namely their ability to act as a saturable deterministic single 
photon absorber when subjected to inhomogeneous dephasing. This goes beyond the known collective 
enhancement of their coupling to light fields and their ability to emit single photons in a directed fashion 
and provides a strong motivation to realize quantum devices based on an excited state blockade in a 
scalable arrangements.

{\it Acknowledgment:}
Support by the Deutsche Forschungsgemeinschaft (DFG) within SFB / TRR 21
and National Science Foundation under Grant No. NSF PHY05-51164 is acknowledged. 
H. W. is supported by the German Academic Exchange
Service (DAAD) and by the National Science Foundation
through a grant for the Institute for Theoretical Atomic,
Molecular and Optical Physics at Harvard University and
Smithsonian Astrophysical Observatory.



\end{document}